\documentclass{article}

\usepackage{microtype}
\usepackage{graphicx}
\usepackage{subfigure}
\usepackage{booktabs} 
\usepackage{xspace}
\usepackage{xcolor}
\usepackage{amsmath}
\usepackage{amssymb}
\usepackage{enumitem}
\usepackage{booktabs} 
\usepackage{multirow} 
\usepackage{soul}
\usepackage[square,numbers]{natbib}
\usepackage{authblk}
\PassOptionsToPackage{hyphens}{url}\usepackage{hyperref}

\newcommand{\sys}{ThunderServe\xspace}
\newcommand\semismall{\fontsize{8.4}{9}\selectfont}

\newcommand{\ffc}[1]{{\color{black}{#1}}}
\newcommand{\jyh}[1]{{\color{black}{#1}}}
\newcommand{\xzyao}[1]{{\color{black}{#1}}}

\DeclareMathOperator*{\argmax}{arg\,max}

\newcommand{\jyhh}[1]{{\color{black}{#1}}}

\newcommand{\specialcell}[2][c]{%
	\begin{tabular}[#1]{@{}c@{}}#2\end{tabular}}
\newcommand{\mytextcircled}[1]{\textcircled{\raisebox{-0.8pt}{#1}}}

\usepackage{hyperref}

\usepackage[accepted]{mlsys2025}

\begin{document}

\twocolumn[
\mlsystitle{\sys: High-performance and Cost-efficient LLM Serving in Cloud Environments}



\mlsyssetsymbol{equal}{*}

\begin{mlsysauthorlist}
\mlsysauthor{Youhe Jiang}{equal,cam}
\mlsysauthor{Fangcheng Fu}{equal,pku}
\mlsysauthor{Xiaozhe Yao}{equal,eth}
\mlsysauthor{Taiyi Wang}{cam}
\mlsysauthor{Bin Cui}{pku}
\mlsysauthor{Ana Klimovic}{eth}
\mlsysauthor{Eiko Yoneki}{cam}
\end{mlsysauthorlist}

\mlsysaffiliation{cam}{Department of Computer Science, University of Cambridge, Cambridgeshire, UK}
\mlsysaffiliation{pku}{Department of Computer Science, Peking University, Beijing, China}
\mlsysaffiliation{eth}{Department of Computer Science, ETH Zurich, Zurich, Switzerland}

\mlsyscorrespondingauthor{Eiko Yoneki}{eiko.yoneki@cl.cam.ac.uk}

\mlsyskeywords{Machine Learning, MLSys}

\vskip 0.3in

\begin{abstract}
Recent developments in large language models (LLMs) have demonstrated their remarkable proficiency in a range of tasks. Compared to in-house homogeneous GPU clusters, deploying LLMs in cloud environments with diverse types of GPUs is crucial for addressing the GPU shortage problem and being more cost-effective. However, the diversity of network environments and various GPU types on the cloud bring difficulties to achieving high-performance serving. In this work, we propose \sys, a high-performance and cost-efficient LLM serving system for heterogeneous cloud environments. We introduce a \textit{novel scheduling algorithm}, which optimizes the deployment plan of LLM serving to accommodate the heterogeneous resource and network bandwidth conditions in cloud environments. Furthermore, we propose a \textit{lightweight re-scheduling} mechanism, designed to adapt to fluctuating online conditions (e.g., node failures, workload shifts) without the need for costly restarts of ongoing services. Empirical results in both heterogeneous cloud and homogeneous in-house environments reveal that \sys delivers \jyhh{up to a 2.1$\times$ and on average a $1.7\times$} increase in throughput and achieves \jyhh{up to a 2.5$\times$ and on average a $1.5\times$} reduction in latency deadlines compared with state-of-the-art systems given the same price budget, suggesting opting for cloud services provides a more cost-efficient solution.
\end{abstract}
]



\printAffiliationsAndNotice{\mlsysEqualContribution} 

\section{Introduction}
\label{sec:1_intro}

Large Language Models (LLMs) such as GPT \cite{achiam2023gpt}, LLaMA \cite{touvron2023llama}, OPT \cite{zhang2022opt} and Falcon \cite{falcon180b} have demonstrated strong performance across a wide range of advanced applications. However, serving LLMs is cost-demanding, requiring a large amount of hardware accelerators like GPUs to satisfy efficiency requirements such as latency and throughput.

Mainstream LLM serving systems primarily focus on high-performance GPUs like NVIDIA A100 and H100 in homogeneous GPU clusters. 
However, it is difficult for many LLM service providers to get access to sufficient high-performance GPUs, either due to the well known GPU shortage problem \cite{skypilot,euromlsysworkshop} or the substantial fees. 
Meanwhile, with more and more advanced GPU architectures announced in the past few years, there are many less-performant GPUs in former generations remaining under-utilized.
Thus, real-world cloud environments usually consist of heterogeneous GPUs and diverse prices. As shown in Table~\ref{tab:gpu}, cloud environments offer a wide range of hardware specifications and rental prices, providing users with diverse options to reduce the costs associated with LLM deployment and serving. Recent efforts \cite{jiang2024hexgen,mei2024helix,griggs2024melange,miao2023spotserve} have demonstrated that serving LLM with heterogeneous GPUs presents opportunities in reducing the serving cost.
%
However, we find that these heterogeneous serving systems mainly address the heterogeneity in \textit{hardwares} but fail to take account of the heterogeneity in \textit{the computation and memory-access workloads} of different inference phases\xzyao{, which hinders the utilization of GPU resources}. 
Such heterogeneity mainly comes from the distinct characteristics of LLM inference, and has raised a surge of research interests.

\begin{table}[!t]
\centering
\caption{\small{GPU specifications and pricing}}
\small
\resizebox{\linewidth}{!}{
\begin{tabular}{l|c|c|c|c}
\hline
\specialcell{\textbf{GPU}\\\textbf{Type}} & \specialcell{\textbf{Memory Access}\\\textbf{Bandwidth}} & \specialcell{\textbf{Peak}\\\textbf{FP16 FLOPS}} & \specialcell{\textbf{Memory}\\\textbf{Limite}} & \specialcell{\textbf{Price}\\\textbf{(per GPU)}} \\
\hline
A100 & 2 TB/s & 312 TFLOPS & 80 GB & \$1.753/hr \\
A6000 & 768 GB/s & 38.7 TFLOPS & 48 GB & \$0.483/hr \\
A5000 & 626.8 GB/s & 27.8 TFLOPS & 24 GB & \$0.223/hr \\
A40 & 696 GB/s & 149.7 TFLOPS & 48 GB & \$0.403/hr \\
3090Ti & 1008 GB/s & 71 TFLOPS & 24 GB & \$0.307/hr \\
\hline
\end{tabular}
}
\label{tab:gpu}
\vspace{-1em}
\end{table}

Recent works~\cite{patel2023splitwise,hu2024inference,qin2024mooncake,jin2024p} have designed phase splitting approaches to utilize different amount of computational resources for prefill and decoding phase in LLM inference, which involves partitioning the two phases onto separate devices and transmitting the intermediate results (primarily KV caches) between them.
Many empirical evidences have shown that such phase splitting approaches increase overall hardware utilization and system efficiency compared with the phase co-locating counterparts.



\begin{figure}[!t]
  \centering
  \includegraphics[width=0.7\linewidth]{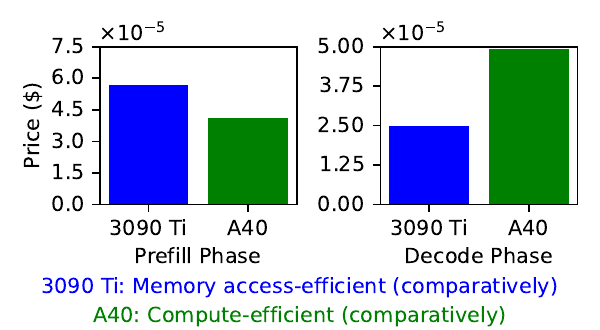} 
  \vspace{-1em}
  \caption{\small{Prefill and decode prices for a single request with input and output lengths of 512 and 16 on 3090Ti and A40.}}
  \label{fig:a40}
  \vspace{-1em}
\end{figure}

As the heterogeneity exists in both hardwares and \xzyao{workload characteristics (i.e., compute/memory-bound)}, we suggest that \textit{the phase splitting approach fits the heterogeneous capabilities among GPUs in cloud environments well}. 
In particular, since the two phases differ in the workload characteristics, it is an intuitive idea to leverage different types of GPUs for the two phases. For instance, as illustrated in \autoref{fig:a40}, the A40 GPU with 149.7 TFLOPS is more cost-effective for the compute-intensive prefill phase, whereas the 3090Ti with 1008 GB/s memory bandwidth is better suited for the memory-bounded decode phase.
Inspired by this, this work presents the first effort to integrate the phase splitting idea with the heterogeneity among GPUs, aiming to achieve high-performance and cost-effective LLM serving in cloud environments.
Nevertheless, the unique attributes of cloud environments pose three key challenges:



\noindent \textbf{{Challenge 1: heterogeneous and limited resource pool.}}
The available GPUs in cloud environments are usually in heterogeneous types, each with distinct specification (e.g., peak FLOPS, device memory bandwidth, and device memory limit), and the amount of each type is also restricted~\cite{skypilot,euromlsysworkshop}.
As a result, to deploy multiple copies (a.k.a. model replicas) of the same LLM, we must consider how to organize the available GPUs from a global view --- given the available resources of diverse types, we need to jointly consider which GPUs should be grouped together to serve one model replica, and whether this replica should serve as the prefill or decoding phase. To our knowledge, this is an unexplored problem so far.

\noindent \textbf{{Challenge 2: heterogeneous and low network bandwidth.}}
The second essential characteristic of cloud environments is that GPUs are usually connected through low network bandwidth, typically, PCIe for intra-node and ethernet for inter-node communication. And the network bandwidth also exhibits a high level of heterogeneity across different pairs of GPUs due to the discrepancy in connectivity (different PCIe versions, node locality, etc.). 
Such a network condition raises a hurdle for efficient LLM serving.
On the one hand, transmitting KV caches from prefill to decode replicas inevitably incurs significant communication volume. 
While prior works~\cite{patel2023splitwise,zhong2024distserve} simply assume high-speed network connections (e.g., NVLink and Infini-Band) are available and overlook the communication overhead of KV caches (detailed in \S\ref{subsec:distributedllmdeployment}), which is impractical for clouds.
On the other hand, due to the astonishing size of LLMs, model parallelism has been a cornerstone for LLM deployment. Thus, there expresses a need for designing heterogeneity-aware parallelization to facilitate phase splitting LLM serving on clouds. 

\noindent \textbf{{Challenge 3: workload variability.}}
Compared to in-house clusters, resources on clouds are more unstable \cite{miao2023spotserve,duan2024parcae,yousif2018cloud,erben2024can}, and the distribution of requests (e.g., average arrival rate, input and output length) may change over time in online services in practice \cite{wang2024burstgpt}.
These factors exacerbate the variability of serving workloads in cloud environments.
In order to adapt to such workload variations, prior works~\cite{zhong2024distserve} necessitate two steps: re-generating the deployment plan from scratch and re-loading the LLM parameters to adjust the model deployment.
However, both steps are costly.
Re-generating the deployment plan could take minutes to complete due to the complex hardware environments on cloud, and re-loading the LLM with a huge amount of parameters could be time-consuming. For instance, loading a 175B model with a disk bandwidth of 1.2 GBps takes over five minutes.
Such expensive steps would lead to severe interruption to the online services.



%
 
To address these challenges, we develop \sys, an efficient and robust LLM serving system on clouds. 
\ffc{Our contributions are summarized as follows:}

\textbf{Contribution 1:} We formulate the scheduling problem of LLM deployment and serving on cloud as a two-level hierarchical optimization problem, and develop a novel scheduling algorithm to optimize the deployment plan. 
In the upper-level, we develop a tabu search algorithm to partition the available GPUs of diverse types into model serving groups (with each group responsible for one model replica). In the lower-level, we determine the optimal parallel configuration for each group as well as the orchestration of prefill and decode replicas to optimize GPU and network usage.


\textbf{Contribution 2:} We design a lightweight re-scheduling mechanism, which only involves adjusting the phase designation and orchestration in real-time, accelerates the re-generation of deployment plan by a large extent, and does not need to re-load the LLM parameters. It enables our system to adapt to workload shifts at minimal cost, thereby enhancing the robustness of LLM serving on cloud.

\textbf{Contribution 3:} Based on these techniques, we implement \sys, an efficient LLM serving system for clouds featuring phase splitting. \sys allows the two phases of LLM inference to be split onto separate GPUs with different resource allocations and parallel strategies. We further integrate a KV cache compression technique into our system, which performs a one-shot compression on the KV cache for efficient inter-phase communication on clouds while maintaining the model quality.

\textbf{Contribution 4:} \xzyao{The performance of \sys in the cloud environment is evaluated through comprehensive experiments. We compare its system and economic efficiency with state-of-the-art LLM serving systems, including HexGen in the same heterogeneous cloud environment, as well as DistServe and vLLM in a homogeneous in-house setting given the \textit{same} budget in terms of cloud service fees}. The empirical results demonstrate that \sys achieves \jyhh{up to 2.1$\times$ and on average 1.7$\times$} increase in throughput and \jyhh{up to 2.5$\times$ and on average 1.5$\times$} reduction in latency compared with existing systems, showcasing the potential of cost-effective LLM serving over clouds.

\section{Background and Related Works}
\label{sec:2}

\begin{figure}[!t] 
  \centering
  \includegraphics[width=0.8\linewidth]{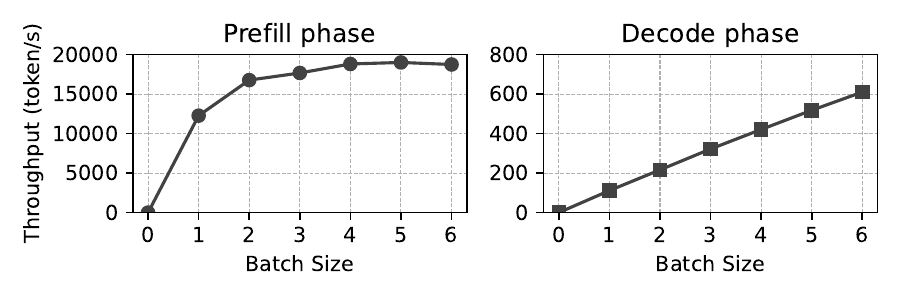} 
  \vspace{-1em}
  \caption{\small{Effects of batching on different phases (LLaMA-7B with each input having a sequence length of 1024).}}
  \label{fig:prefillvsdecode}
  \vspace{-1em}
\end{figure}

\noindent \textbf{Phases of LLM inference.}
Given an input prompt, the inference process of LLMs typically consists of two phases: \textit{the prefill phase} processes the prompt to compute the key-value (KV) cache and generates the first token in a single step, and \textit{the decode phase} takes the last generated token and KV cache as inputs to generate subsequent tokens. Different from the prefill phase, the decode phase is executed for several steps, with each step generating only one token, which makes the decode phase more memory bandwidth bounded than the computationally intensive prefill phase. 

\noindent \textbf{Performance metrics.}
There are three key metrics to evaluate LLM inference: \textit{time to first token (TTFT)}, which measures the time of generating the first token; \textit{time per output token (TPOT)}, which quantifies the average time of generating each token in the decode phase; \textit{end-to-end (E2E) latency}, which assesses the overall processing time of a request (includes queuing, prefill, and decode costs).
For LLM serving systems, there are also two key metrics: \textit{Service Level Objective (SLO) attainment}, which represents the percentage (e.g., 99\%) of requests that can be served within a predefined time frame set by the SLO, and the SLO is often scaled to different multiples of single device execution latency (denoted as SLO scale) to evaluate system performance under different levels of SLO stringency; \textit{throughput}, which measures the number of requests a system can handle within a specified time period. Efficient LLM serving systems should optimize either of the metrics, and meanwhile meet the performance requirements of specific applications if necessary.


\noindent \textbf{Batching.}
\label{sec:batching}
Due to the divergent workloads of the two phases, integrating batching strategies results in performance variations. 
%
In the prefill phase, a small batch size quickly saturates the GPU, yielding marginal benefits from further batching. Besides, execution latency increases linearly with batch size, rendering batching impractical when the TTFT constraints are strict. 
As shown in \autoref{fig:prefillvsdecode}, we conduct a small testbed with LLaMA-7B as an example, which reveals that when the total number of tokens in a batch exceeds 1024, GPU efficiency reaches a plateau rather than being further enhanced, showcasing the limited effect of batching on system performance during the prefill phase. 
%
Due to the token-by-token processing nature of the decode phase, batching is essential for preventing low GPU utilization and enhancing efficiency. As shown in \autoref{fig:prefillvsdecode}, increasing the batch size enhances GPU efficiency\xzyao{, particularly in the decode phase}. In short, batching in the decode phase improves performance, even under stringent TPOT constraints.

\label{subsec:distributedllmdeployment}

\noindent \textbf{Parallelism strategies.}
To parallelize the model over multiple GPUs, there are two prevalent forms of model parallelism, which are tensor model parallelism and pipeline model parallelism.
\textit{Tensor model parallelism (TP)} \cite{shoeybi2019megatron,nagrecha2021model} divides model weights and computationally intensive operations such as matrix multiplication across various GPUs, thereby splitting data scanning and computation to minimize LLM inference latency, particularly the TTFT in the prefill phase. 
\textit{Pipeline model parallelism (PP)} \cite{huang2019gpipe,narayanan2019pipedream} divides the layers of a model into multiple stages. These stages are assigned to distinct GPUs for execution and they establish a pipeline. Only inter-layer activations are needed to be communicated between stages. 

\noindent \textbf{Phase splitting deployment.}
As the prefill and decode phases differ in \xzyao{workload characteristics (i.e., compute/memory-bound) significantly}, recent efforts propose to utilize different hardware resources for the two phases in order to avoid performance interference \cite{patel2023splitwise,zhong2024distserve,hu2024inference,jin2024p,qin2024mooncake}. In other words, there are two kinds of model replicas, one for prefill and the other for decode, respectively, and it is necessary to transmit the KV cache from the prefill replicas to the decode ones. 
Due to the substantial size of KV cache, existing efforts essentially require high communication bandwidth for the KV cache transfer ---
\citet{zhong2024distserve} proposed to colocate prefill and decode replicas on GPUs within the same node, facilitating fast KV cache transfer with NVLink, while \citet{patel2023splitwise} and \citet{hu2024inference} utilized high-speed Infini-Band connections for inter-node communication. 
However, in cloud environments, GPUs are connected via limited bandwidth (typically, PCIe for intra-node connections and Ethernet for inter-node connections) rather than high-speed connections (typically, NVLink and Infini-Band). Therefore, KV cache transfer would lead to a huge cost and it necessitates enhancement.

\textbf{Heterogeneous GPU Computing.} Recent research has investigated diverse approaches to deploy large models on heterogeneous GPU clusters. 
HexGen \cite{jiang2024hexgen} proposes asymmetric partitioning and advanced scheduling to deploy generative inference in a decentralized and heterogeneous setting. 
Helix \cite{mei2024helix} formulates the heterogeneous GPUs and network connections as a maxflow problem and adopts a mixed integer linear programming algorithm to discover highly optimized strategies to serve LLMs. 
Our work has a similar objective, but is the first effort that integrates phase splitting with the heterogeneous GPUs to provide high-performance cloud serving for LLMs.

\section{Scheduling in \sys}
\label{sec:scheduling}
This section \xzyao{introduces} our scheduling algorithm, which aims to optimize the overall \jyhh{SLO attainment} of the serving. 

\begin{figure}[!t] 
  \centering
  \includegraphics[width=\linewidth]{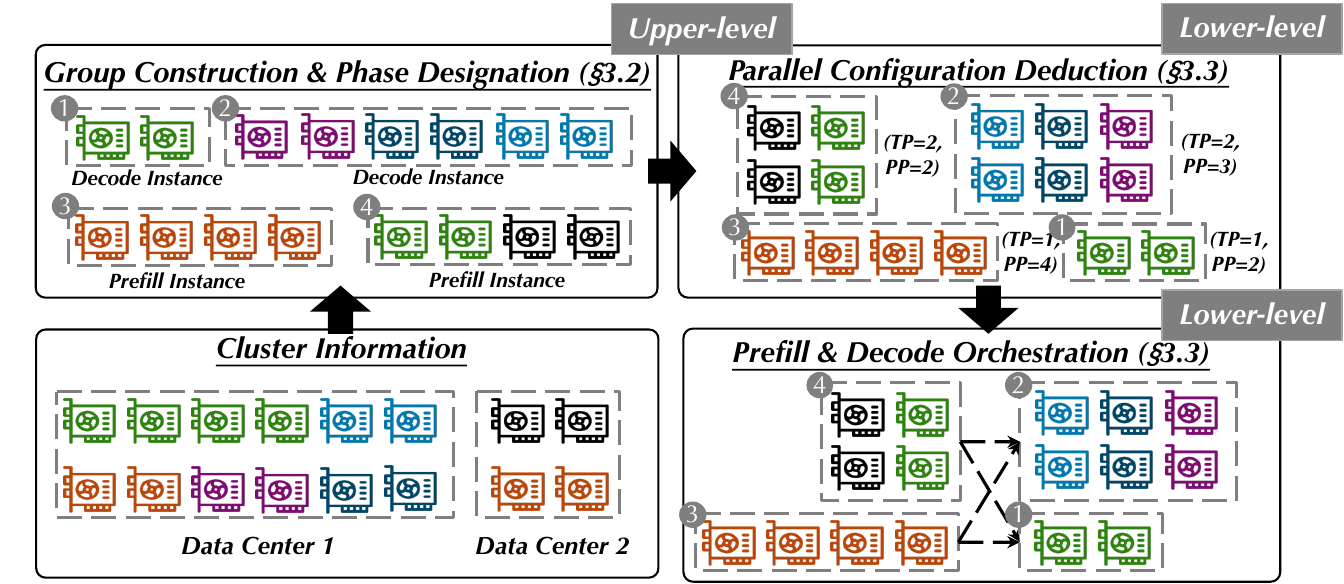} 
  \vspace{-1em}
  \caption{\jyh{\small{Workflow of our scheduling algorithm.}}}
  \vspace{-1em}
  \label{fig:scheduling_routine}
\end{figure}

\subsection{Overview and Problem Formulation}

To describe the model deployment over the available resources of heterogeneous capabilities, the scheduling algorithm should produce four essential components: \mytextcircled{1} The \ul{\textit{group construction}}, i.e., how to partition the GPUs into multiple model serving groups, where each group is responsible for one model replica. 
\mytextcircled{2} The \ul{\textit{phase designation}} that indicates whether each group should serve as the prefill or decode phase. 
\mytextcircled{3} The \ul{\textit{parallel configuration}} for each model replica.
\mytextcircled{4} The \ul{\textit{orchestration of prefill and decode replicas}} to guide how the requests should be routed. We term a solution to these four components as a \textbf{\ul{\textit{deployment plan}}}.

As each deployment plan consists of four components, there is an extremely huge solution space. To ease the scheduling, we decouple the huge solution space into the Cartesian product of two sub spaces, by turning the derivation of deployment plan into a two-level hierarchical optimization problem as follows. 
\begin{itemize}[topsep=5pt, leftmargin=*]
 
\item \textbf{Upper-level:} 
Suppose there are $G$ GPUs of $T$ types in total, and $G_{t}$ denotes the number of GPUs of type $t$. The objective of the upper-level problem is to find out the best combination of group construction and phase designation that maximizes the end-to-end \jyhh{SLO attainment.}

\item \textbf{Lower-level.} 
Given the group construction and phase designation, 
the objective of the lower-level problem is to determine the best parallel configuration for each group and how the prefill and decode replicas should be orchestrated to maximize the end-to-end \jyhh{SLO attainment}.
 
\end{itemize} 
Obviously, the Cartesian product of the solution spaces of the two problems completely covers all possible deployment plans, so finding the optimal deployment plan is equivalent to solving the hierarchical optimization problem by nature. 
Figure \ref{fig:scheduling_routine} shows the workflow of our scheduling algorithm. 
Given the target model and the available resources, we initiate the GPU construction and phase designation process, which involves a tabu search process that iteratively proposes solutions to the upper-level problem (\S\ref{sec:schedule_upper_level}). 
Then, for any possible solution to the upper-level problem, we solve the lower-level problem to obtain its performance, which involves the the parallel configuration deduction and phase orchestration process (\S\ref{sec:schedule_lower_level}).

\subsection{Solving the Upper-level Problem}
\label{sec:schedule_upper_level}
In cloud environments, there are various types of GPUs with heterogeneous capabilities, making the group construction a non-trivial problem. In addition, the phase-splitting design requires meticulous phase designation to the model serving groups for better performance, making the upper-level problem even more complex. 
Formally, in \autoref{appendix:a}, we show that the upper-level problem is essentially a job shop scheduling problem (JSSP), which is a notoriously difficult NP-hard problem in combinatorial optimization \cite{sotskov1995np,omar2006job}.

\begin{algorithm}[tb]
\caption{\small{Routine of solving the upper-level problem based on tabu search, where $N_{\text{step}}$ denotes the number of search steps, $N_{\text{nghb}}$ denotes the number of neighbors to navigation in each step, $N_{\text{mem}}$ denotes the maximum number of memorized solutions, and $f(\cdot)$ denotes the performance of a solution evaluated by solving the lower-level problem.}}
\label{alg:tabu}
\small
\begin{algorithmic}[1]
\FUNCTION{\textsc{TabuSearch}($N_{\text{step}}$=100, $N_{\text{nghb}}$=10, $N_{\text{mem}}$=5):}
\STATE Initialize the current solution $x$
\STATE Initialize tabu list $T \gets []$, best solution $x_{best} \gets x$
\STATE {\color{gray}{/* Iterative Neighborhood Search */}}
\FOR{$N_{\text{step}}$ search steps}
    \STATE Construct $N_{\text{nghb}}$ neighbors of $x$ and exclude those in $T$ to form the neighborhood set $\mathcal{N}$ for navigation
    \STATE $x' \gets \argmax_{x'' \in \mathcal{N}} f(x'')$
    \STATE \textbf{if} $f(x') > f(x_{best})$ \textbf{then} $x_{best} \gets x'$
    \STATE $T.\operatorname{append}(x')$
    \STATE \textbf{if} $\operatorname{len}(T) > N_{\text{mem}}$ \textbf{then} $T \gets T[-N_{\text{mem}}:\;]$
    \STATE $x \gets x'$
\ENDFOR
\STATE \textbf{return} $x_{best}$
\ENDFUNCTION
\end{algorithmic}
\end{algorithm}

A well-known approach to solve JSSP is tabu search \cite{glover1990tabu,glover1998tabu,gendreau2005tabu} and there have been many efforts applying tabu search to solve JSSP under various situations \cite{hurink1994tabu,tabu_jssp_1997,zhang2007tabu}. Motivated by this, we adapt tabu search to the upper-level problem and design a brand new algorithm to identify the optimal deployment plan, which is demonstrated in Algorithm \ref{alg:tabu}. 
In essence, it starts from an initial solution, and leverages an iterative neighborhood search process to improve the solution. 
Below we focus on how to determine the initial solution and how to construct neighbors given the current solution in our scenario.

\noindent \textbf{Initialization.} It is essential to have a good initial solution in tabu search in order to speedup the search process and escape from local optima. 
Thus, we utilize the Hierarchical Clustering method \cite{shetty2021hierarchical} to cluster the GPUs according to their inter-connection bandwidth matrix, and subsequently treat each generated cluster as one model serving group at initialization.
Intuitively, this makes the initial assignment of model serving groups strategically avoid connections with ultra-low communication bandwidth in the cloud environment. In addition, the phase designation of each group is randomly initialized.

\begin{figure}[!t]
  \centering
  \includegraphics[width=\linewidth]{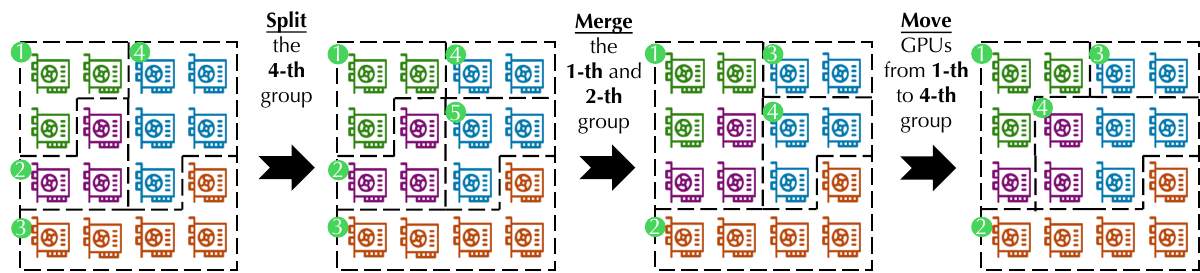} 
  \vspace{-1em}
  \caption{\jyh{\small{Examples of neighbor construction in tabu search (changes in phase designation are omitted for simplicity).}}}
  \label{fig:tabusearch graph}
  \vspace{-1em}
\end{figure}

\noindent \textbf{Neighbor construction.} 
In the iterative search process, tabu search evaluates a set of neighboring solutions given the current solution. Denote $g_{i,t}$ as the number of GPUs of type $t$ in group $i$. We provide four approaches of to construct the neighboring solutions, as exemplified in \autoref{fig:tabusearch graph} and detailed below.
\begin{itemize}[leftmargin=*,topsep=5pt]
\item \textit{Flipping phase designation.} This approach randomly selects a group and flips its phase. In other word, if the group is originally designated to serve as a prefill replica, then it will be changed to a decode replica, and vice versa. 
\item \textit{Splitting a group into two.} This approach randomly selects and splits a group {\semismall \( g_{s,t} \)} into two based on a random ratio {\semismall \( r \)}, assigning {\semismall \( \lfloor g_{s,t} \times r \rfloor \)} GPUs to the first new group {\semismall \( g_{s_1,t}' \)} and the remainder to the second group {\semismall \( g_{s_2,t}' \)}, which is effective for exploring how dividing resources impacts performance, particularly when a group might be overly large or tasked beyond its efficient operating capacity, i.e.,
\begin{equation*}
g_{s_1,t}' \leftarrow \left\lfloor g_{s,t} \times r \right\rfloor, 
g_{s_2,t}' \leftarrow g_{s,t} - g_{s_1,t}', \forall t \in \{1, \ldots, T\}.
\end{equation*}
The phase of the new groups will be randomly designated.
\item \textit{Merging two groups into one.} This approach randomly selects and merges two groups {\semismall \( g_{i,t}, g_{j,t} \)} into one, which explores the potential benefits or drawbacks of resource centralization for individual model serving, i.e.,
\begin{equation*}
g_{\text{merged},t}' \leftarrow g_{i,t} + g_{j,t}, \quad \forall t \in \{1, \ldots, T\}.
\end{equation*}
The phase of the new group will be randomly designated.
\item \textit{Moving GPUs between groups.} This approach involves moving a certain number of (denoted as {\semismall \( m_t \)}) GPUs of type {\semismall \( t \)} from group {\semismall \( i \)} to group {\semismall \( j \)}. 
It is useful for exploring the effects of resource reallocation in scenarios where different groups may benefit from different GPU capabilities.
\begin{equation*}
g_{i,t}' \leftarrow g_{i,t} - m_t, \quad g_{j,t}' \leftarrow g_{j,t} + m_t.
\end{equation*}
\end{itemize}
The adjustment in group construction and phase designation iteratively navigate the neighborhood space of the current solution, enabling tabu search to explore potential performance enhancements.
Additionally, to expedite the search process, early checks are performed for each generated neighborhood. For instance, if the total GPU memory of any serving group after the moving or splitting operations is insufficient to hold even a single copy of the model parameters, then the constructed neighbor is eliminated from further evaluation.

\subsection{Solving the Lower-level Problem}
\label{sec:schedule_lower_level}
The goal of the lower-level problem is two-fold, i.e., finding the optimal parallel configuration for each model serving group and how to orchestrate all groups together. Fortunately, since the available resources and the designated phase of each group are given, we find that the deduction of optimal parallel configuration is independent to the orchestration. To be specific, suppose $C'$ is a parallel configuration for an arbitrary group, and $C''$ is another configuration with higher performance, then it is obvious that we can always find out an orchestration with $C''$ that is at least as good as that with $C'$. As a result, we first deduce the optimal parallel configuration for each group individually, and then determine the orchestration, as introduced below.

\noindent \textbf{Deduction of parallel configuration.} Given the available resources and the designated phase of each group, we wish to deduce the optimal parallel configuration. 
As discussed in \S\ref{sec:batching}, the two phases differ in workload characteristics, so their desirable parallel configurations also vary. 
For groups that serve as prefill replicas, we aim to deduce the latency-optimal parallel configurations, since the prefill phase is computation-intensive and batching does not help to enhance efficiency.
In contrast, for groups that serve as decode replicas, we aim to deduce the throughput-optimal parallel configurations, since this memory bandwidth bounded phase benefits from batching. 

Numerous studies \cite{jiang2024hexgen,zheng2022alpa,li2023alpaserve,Miao_2022,miao2023spotserve,wang2024improving} have investigated how to deduce the optimal parallel configuration by meticulously enumerating a vast number of possible configurations. In this work, we further take the key characteristics of cloud environments into account and design heuristics to accelerate the enumeration process. We introduce the heuristics below and leave the details of the deduction routine in \autoref{appendix:b}. 

\begin{itemize}[topsep=5pt, leftmargin=*]
     
    \item Typically, cloud services generally do not provide rapid links among nodes. Thus, we disallow tensor model parallelism to be deployed over GPUs across different nodes due to its demand of higher network bandwidth. 

    \item Since different types of GPUs may differ in the memory capacity and computing ability, we support non-uniform pipeline layer partitioning for pipeline model parallelism. 

    \item In response to the heterogeneity in inter-node communication, we employ a dynamic programming algorithm that aims to identify the path minimizing the cross-stage communication cost in pipeline model parallelism.

\end{itemize}

\begin{figure}[!t] 
  \centering
  \includegraphics[width=0.8\linewidth]{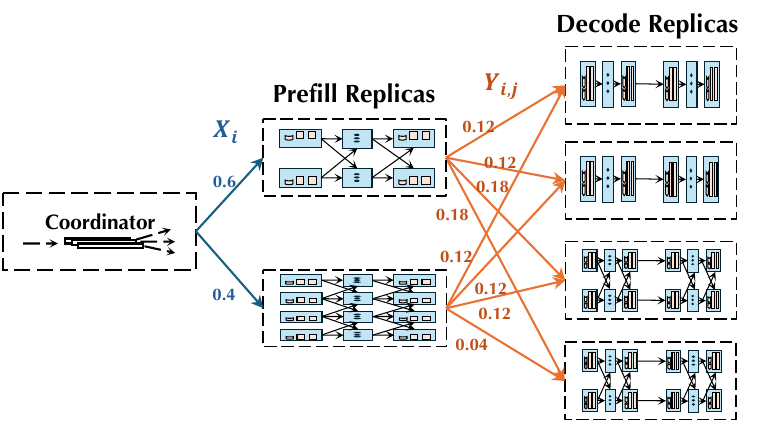} 
  \caption{\small{An example orchestration of prefill and decode replicas.}}
  \label{fig:orches}
  \vspace{-1em}
\end{figure}

\noindent \textbf{Orchestration of prefill and decode replicas.}
Due to the network heterogeneity in cloud environments, it is essential to identify the optimal orchestration of prefill and decode replicas within the cluster to minimize KV cache communication cost and optimize overall \jyhh{SLO attainment}.

\jyhh{We adopt the inference task simulator from DistServe \cite{zhong2024distserve}}, which estimates the \jyhh{SLO attainment} according to workload information (e.g., input length, output length, etc.) and single request processing time. It is noteworthy \jyhh{that we integrate the KV cache communication cost into the simulator,} since it is non-negligible in cloud environments. To be specific, suppose there are $m$ prefill replicas and $n$ decode replicas, our simulator enumerates every pair of them and estimates the \jyhh{SLO attainment} by integrating the KV cache communication cost, which is analyzed via the alpha-beta model \cite{hockney1994communication}:
\begin{equation}
\semismall
\label{eq:kv_comm_cost}
  {T}^{(kv\_comm)}_{ij} = {\alpha_{ij}} + {2bsh\text{N}_{\text{bytes}}}/{\beta_{ij}},
\end{equation}
where \( b, s \) represent the batch size and sequence length for inference, \( h \) represents the hidden size of a Transformer block, \(\text{N}_{\text{bytes}} \) represents the byte size for KV cache communication, and \( \alpha_{ij}, \beta_{ij} \) represent the network latency and bandwidth between the $i$-th prefill replica and the $j$-th decode replica. \jyhh{We evaluate the simulator and alpha-beta model accuracy in \autoref{appendix:simu}.}

Based on this, we estimate the \jyhh{SLO attainment} of every pair of prefill and decode replicas. Formally, denote $D \in \mathbb{R}^{m \times n}$ as the \jyhh{SLO attainment} matrix, where $D_{ij}$ represents the estimated \jyhh{SLO attainment} when requests are processed by the $i$-th prefill replica and the $j$-th decode replica. Then, we turn the optimization problem of overall system \jyhh{SLO attainment} into a simple two-stage transportation problem (TSTP)~\cite{santoso2022development}
as follows:
\begin{equation*}
\semismall
\begin{aligned}
& \argmax_{X \in \mathbb{R}^{m}, Y \in \mathbb{R}^{m \times n}} 
\sum_{i=1}^{m} \sum_{j=1}^{n} X_i Y_{ij} D_{ij} \\
& \text{s.t. } 
\sum_{i=1}^{m} X_i = 1, \;
\sum_{j=1}^{n} Y_{ij} \text{ for }\forall i, \;
X_i \geq 0 \text{ for }\forall i, \;
Y_{ij} \geq 0 \text{ for }\forall i, j,
\end{aligned}
\end{equation*}
where $X_i$ denotes the portion of incoming requests that are assigned to the $i$-th prefill replica, and $Y_{ij}$ denotes the portion of requests processed by the $i$-th prefill replica that are dispatched to the $j$-th decode replica. The TSTP can be solved by linear programming, and the optimal $X^*, Y^*$ describe how requests are routed among the model serving groups, which also represent the orchestration of different replicas to maximize the overall system \jyhh{SLO attainment}.

By combining the deduction of optimal parallel configuration and the orchestration of different replicas, we accomplish the solution to the lower-level problem, and the resulting system \jyhh{SLO attainment} is returned to the tabu search process (i.e., $f(\cdot)$ in Algorithm \ref{alg:tabu}).


\begin{figure}
    \centering
    \includegraphics[width=\linewidth]{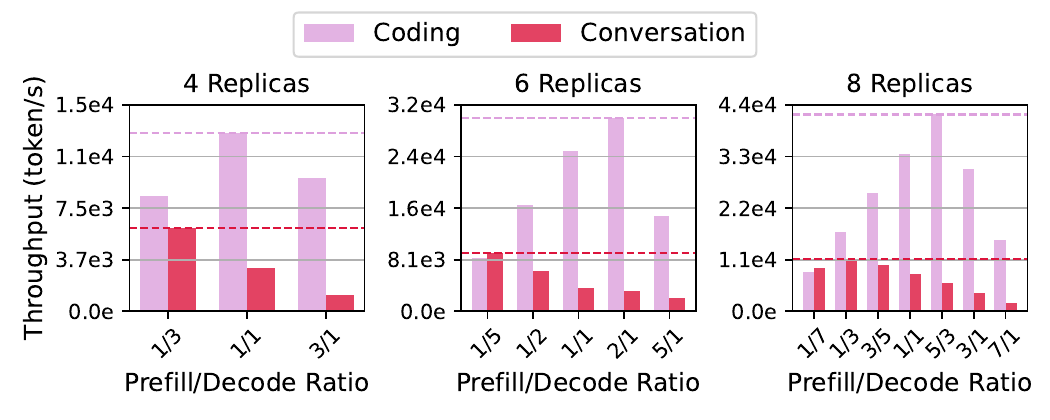}
    \vspace{-2em}
    \caption{\small{Throughput (token/s) by prefill-to-decode ratio. Impact of phase designation and orchestration on overall system throughput. We experiment with LLaMA-13B on both coding and conversation workloads across clusters with 8, 12, and 16 A5000 GPUs, respectively, with two GPUs serving one replica. The ratio represents the prefill-to-decode ratio (i.e., the ratio of \# prefill replicas to \# decode replicas). We have also provided SLO Attainment results in \autoref{fig:ratio_goodput} of \autoref{appendix:ratio}.}}
    \label{tab:throughputvsratio}
    \vspace{-1em}
\end{figure}

\subsection{Lightweight Rescheduling in Real Time}
\label{sec:method_light_reschedule}
Numerous factors in cloud services affect the optimal deployment plan, with two primary factors being LLM inference workloads and GPU availability. On the one hand, LLM services usually exhibit significant variation in workload characteristics across different downstream tasks. For instance, coding workloads typically generate shorter responses than conversational workloads but usually have longer prompts \cite{patel2023splitwise}. 
On the other hand, compared to in-house clusters, cloud resources are inherently more dynamic and unstable, necessitating a good support for cluster size adjustments in real-time. 
Consequently, rescheduling is essential for \sys to adapt the deployment plan to varying online workloads and cluster size changes on cloud.
However, altering the deployment plan is far from trivial in cloud environments. If we re-run the scheduling algorithm from scratch and reload the model parameters according to the updated deployment plan (take minutes to complete), it would lead to severe interruption to the online services. 
Therefore, 
we propose a \textit{lightweight rescheduling} process that only adjusts the phase designation and orchestration in the deployment plan to accommodate varying workloads and cluster sizes.

The rationality behind our lightweight rescheduling is that the changes in workload generally influence the demands on the prefill and decode phases. To elaborate, we conduct comprehensive experiments to demonstrate the impact of phase designation and orchestration over diverse workloads and cluster sizes (with fixed group construction and parallel configuration). 
As shown in \autoref{tab:throughputvsratio}, the coding workload, characterized by relatively longer input length and shorter output length, exhibits enhanced performance with more prefill replicas and fewer decode replicas. Conversely, the conversation workload, typified by relatively shorter prompts and longer responses, necessitates more decode replicas and fewer prefill replicas to prioritize resources to the long-running decoding, with the ideal prefill-to-decode ratio fluctuating as the cluster size varies. 
These findings underscore the critical importance of precise adjustments in the phase designation and orchestration to achieve optimal system performance and realize the ability to adapt to various workloads and cluster sizes.

The lightweight rescheduling is done by simplifying the routines introduced in \S\ref{sec:schedule_upper_level} and \S\ref{sec:schedule_lower_level}: 
\begin{itemize}[topsep=0pt, leftmargin=*]
\item For the tabu search process, only the flipping phase designation approach is used to construct neighboring solutions, while the other approaches are not involved in the lightweight rescheduling.
\item The deduction of parallel configuration is skipped and the orchestration problem will be solved using the unaltered parallel configurations and the newly designated phases.
\end{itemize}
Although our lightweight rescheduling leads to sub-optimality, our experiments in \S\ref{sec:effectiveness_and_ablation} show that it achieves comparable performance against rescheduling from scratch in various scenarios. More importantly, by merely adjusting the phase designation and orchestration, there is no need to reload the parameters, and thus introduces almost zero overhead to the online services. Consequently, our lightweight rescheduling improves the flexibility and robustness of \sys to a great extent.
\section{Implementation}
\label{sec:impl}

\sys is a distributed LLM serving system designed to optimize online services in cloud environments, which develops a novel scheduling algorithm to partition the given cloud GPU resources into model serving groups, designate which phase each group should serve as, deduce the optimal parallel configuration for each group, and determine how the requests should be routed among groups.
It is implemented using 20K lines of Python and C++/CUDA code. \ffc{Besides, \sys incorporates FlashAttention \cite{dao2022flashattention} and PagedAttention \cite{kwon2023efficient} to accelerate LLM inference, and leverages the batching strategy proposed by \citet{zhong2024distserve} for LLM serving.}


\begin{table}[!t]
\centering
\small
\caption{\jyh{\small{Impact of KV cache communication compression on the model accuracy on CoQA, TruthfulQA and GSM8K tasks.}}}
\jyh{
\begin{tabular}{c c c c}
\hline
\textbf{Task} &  & \textbf{LLaMA-7B} & \textbf{LLaMA-13B} \\ \hline
\multirow{2}{*}{CoQA} & 16-bit & 63.95 & 66.35 \\ 
                                    & 4-bit  & 64.58 & 66.54 \\ \hline
\multirow{2}{*}{TruthfulQA} & 16-bit & 30.64 & 29.68 \\ 
                                       & 4-bit  & 30.13 & 29.34 \\ \hline
\multirow{2}{*}{GSM8K} & 16-bit & 13.23 & 22.34 \\ 
                                               & 4-bit  & 12.54  & 21.29  \\ \hline
\end{tabular}
}
\label{tab:acc}
  \vspace{-1em}
\end{table}

\textbf{Overall routine.} The overall routine of \sys is as follows.
\mytextcircled{1} To launch a serving process, the scheduling algorithm (\S\ref{sec:schedule_upper_level} and \S\ref{sec:schedule_lower_level}) generates the deployment plan, which is then utilized to instantiate the model replicas over the cloud GPU resources. 
\mytextcircled{2} During the serving process, the incoming requests are dispatched across the prefill and decode replicas, and the generated responses are gathered.
\mytextcircled{3} At the same time, the inference workload is constantly monitored and reported to the scheduling algorithm.
\mytextcircled{4} Once a workload shift is detected, the scheduling algorithm triggers the lightweight re-scheduling process (\S\ref{sec:method_light_reschedule}) to adjust the deployment plan in response to the new workload.
Due to the space constraint, we refer interested readers to \autoref{appendix:components} for more implementation details of \sys.


\textbf{KV cache compression technique.} As discussed in \S\ref{subsec:distributedllmdeployment}, prior works rely on high-bandwidth connections (i.e., NVLINK or InfiniBand) for transferring KV cache in phase splitting deployment, which is impractical in cloud service scenarios characterized by heterogeneous network conditions among GPUs. 
To reduce the KV cache communication cost, we borrow the idea of low-precision quantization from KIVI \cite{kivi2024} to quantize KV cache to fewer bits, so that the size of each element (i.e. \(\text{N}_{\text{bytes}} \) in \autoref{eq:kv_comm_cost}) is shrinked.
However, unlike existing works in the field of KV cache quantization \cite{kivi2024,kang2024gear}, 
\jyhh{our system does not retain low bitwidths when using the KV cache values for computation.}
Specifically, the KV cache values in the prefill replica are quantized and packed for communication, and then immediately unpacked and dequantized after they are received by the decode replica. \ffc{Thus, both the prefill and decode phases are conducted using the 16-bit KV cache values rather than the quantized ones.}
By this means, we can significantly reduce the KV cache communication volume, without harming the model quality.

To elaborate, we conduct a small testbed with LLaMA-7B over two A5000 GPUs, which featured an inter-communication bandwidth of \jyh{40 Gbps} --- significantly lower than that of InfiniBand and NVLink. 
Quantizing the 16-bit elements to \jyh{4-bit} significantly reduces KV cache communication costs \jyh{from 16-30\% to 4-9\%} of the total end-to-end inference costs, drastically improving the performance of the system. 
\jyhh{Besides, we demonstrate the accuracy results of LLaMA-7B and LLaMA-13B models on CoQA, TruthfulQA and GSM8K tasks with both 16-bit and 4-bit KV cache precision levels. As we do not retain low bitwidths when using the KV cache values for computation, our experiments in \autoref{tab:acc} consistently show that the accuracy drop when using 4-bit precision compared to 16-bit precision remains below 2\% across all experimental scenarios, which confirms the validity of our approach. Due to the space constraint, we provide more evaluation results in \autoref{appendix:ppl}, including perplexity (PPL) and ROUGE-1/2/L on the WikiText2, PTB, and CBT datasets, and the end-to-end throughput comparisons between 16-bit and 4-bit precision.}

\section{Evaluation}
\label{sec:eva}

\subsection{Experimental Setup}
\label{sec:eva_setup}
\noindent \textbf{Hardware environments.}
We consider two types of hardware environments with almost the same price budgets.
\begin{itemize}[topsep=0pt, leftmargin=*]
\item \textit{Heterogeneous GPUs on the cloud.} 
We rent GPUs from Vast.ai, a GPU cloud service provider. We rent four types of instances with 32 GPUs in total: two 4$\times$A6000 instances, \jyh{two 4$\times$A5000 instances}, one 8$\times$A40 instance and two 4$\times$3090Ti instances, with a total price of \$13.542/hour to represent the heterogeneous case.
\item \textit{Homogeneous GPUs in a in-house server.} 
For baseline systems that do not support heterogeneous GPUs, we use one in-house server equipped with 8$\times$A100-80GB GPUs. 
According to \autoref{tab:gpu}, renting the same GPUs costs \$14.024/hour, which is close to the aforementioned price budget on the cloud. 
\end{itemize}
The GPU specifications are provided in \autoref{tab:gpu}, while the network bandwidth can be found in \autoref{appendix:comm_matrix}.

\noindent \textbf{Model and workloads.} We deploy the popular open-source LLaMA-30B model across two real-world workloads, coding and conversation, from the Azure Conversation dataset~\cite{patel2023splitwise}.
And we follow prior works \cite{li2023alpaserve,jiang2024hexgen} to generate the inference workload using a Poisson process determined by the request rate, with consecutive requests (inter-arrival times) following an exponential distribution.

\begin{figure*}[!t]
    \centering
    \begin{minipage}{0.48\linewidth}
        \includegraphics[width=\linewidth]{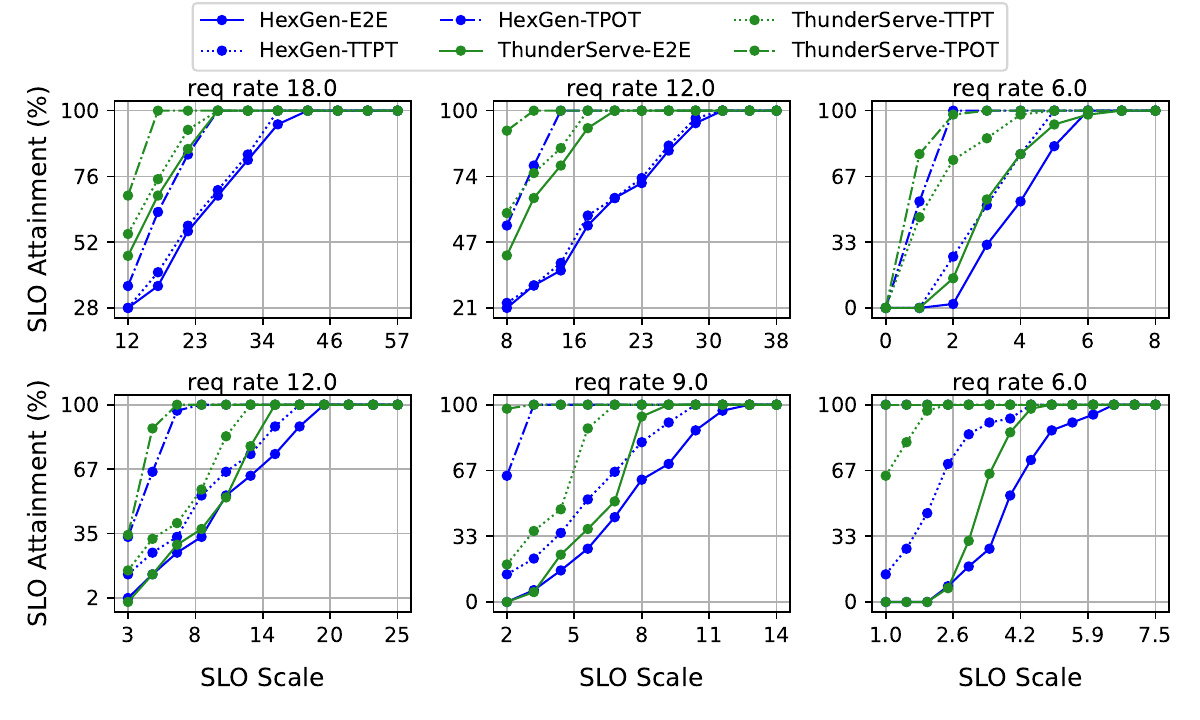}
          \vspace{-2em}
        \caption{\jyh{\small{SLO attainment results on coding (top row) and conversation (bottom row) workloads.}}}
        \label{fig:e2e}
          \vspace{-1em}
    \end{minipage}\hspace{10pt}
    \begin{minipage}{0.48\linewidth}
        \includegraphics[width=\linewidth]{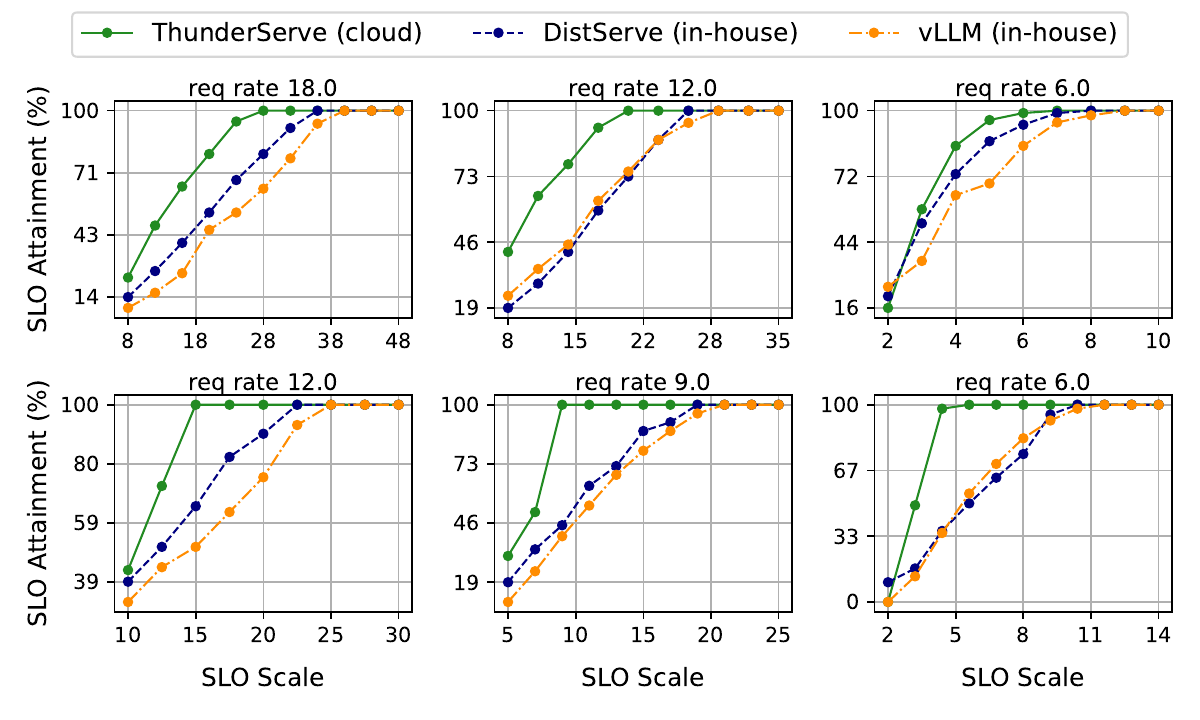}
          \vspace{-2em}
        \caption{\jyh{\small{SLO attainment results on coding (top row) and conversation (bottom row) workloads.}}}
        \label{fig:cloudvsdatacenter}
          \vspace{-1em}
    \end{minipage}
\end{figure*}

\noindent \textbf{Evaluation metrics.} Following prior works \cite{patel2023splitwise,zhong2024distserve}, we focus on overall system SLO attainment and throughput when evaluating the performance. System SLO attainment indicates the percentage of requests completed within a predefined latency deadline. There are three types of SLO: TTFT, TPOT, and E2E SLO. We specifically measure system SLO attainment by the percentage of requests that meet the time criteria established by each SLO type. We scale the SLO to various multiples of the execution latency of A100 GPUs (SLO Scale in \autoref{fig:e2e}), which allows us to evaluate system performance under different levels of operational stringency. For a target \jyhh{SLO attainment} goal (e.g., 90\% and 99\%), we focus on the minimum latency deadline required to achieve the desired attainment.

%

\noindent \textbf{Baselines.} 
We consider state-of-the-art systems under both the cloud and in-house settings, respectively. Under the cloud setting, we consider HexGen \cite{jiang2024hexgen}, an LLM serving system for clouds featuring advanced scheduling and asymmetric parallelism. For the in-house scenario, we consider \ffc{vLLM \cite{kwon2023efficient}, a prestigious LLM serving system, as well as} DistServe \cite{zhong2024distserve}, an LLM serving system featuring phase splitting. 

\begin{figure}[!t]
  \centering
  \includegraphics[width=0.8\linewidth]{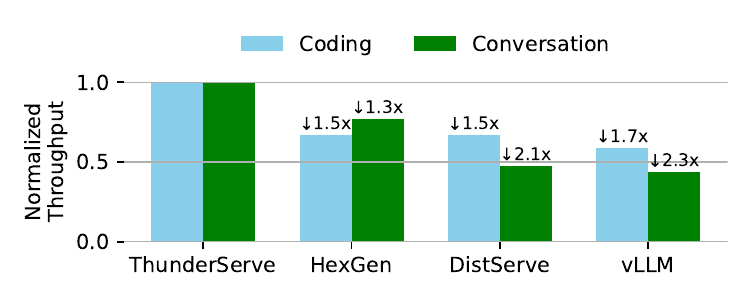} 
    \vspace{-1em}
  \caption{\small{Throughput scaled by \sys.}}
  \label{fig:throughput}
    \vspace{-1em}
\end{figure}

\subsection{End-to-end Evaluation}
In this section, we compare the end-to-end performance of \sys against baselines on various workloads.

\noindent \textbf{System SLO attainment comparisons on the cloud.} 
We first evaluate the performance of \sys on the cloud. As shown in \autoref{fig:e2e}, typically higher average request arrival rates require higher SLO scales (i.e., longer latency deadlines) to meet the \jyhh{SLO attainment goal}. \sys consistently outperforms HexGen in terms of all TTFT, TPOT, and E2E SLO attainments.

On the coding workload, \sys achieves up to 1.8$\times$ and on average 1.4$\times$ lower E2E latency deadlines compared with existing approaches. Specifically, \ffc{as we will elaborate in \S\ref{sec:casestudyofscheduling}}, the coding workload makes our scheduling algorithm designate more prefill replicas than decode replicas, since the bottleneck is on prefilling given the relatively long input prompts. And the prefill-to-decode ratio decreases with the surge of the average request arrival rate, which matches our previous discussion in \S\ref{sec:method_light_reschedule}. On the conversation workload, \sys achieves up to 1.4$\times$ and on average 1.3$\times$ lower E2E latency deadlines. The conversation workload makes our scheduling algorithm deploy more decode replicas than prefill replicas, since the bottleneck is on decoding given the relatively long output responses. The phase splitting technique significantly reduces prefill-decode interference during inference, leading to improved TTFT and TPOT SLO attainments in all cases.

\noindent \textbf{Cost-efficiency of deploying LLM services on the cloud.} To assess the cost-efficiency of deploying LLM services on the cloud, we compare \sys in the cloud setting with DistServe and \jyh{vLLM} in the in-house setting, given the same price budget. As shown in \autoref{fig:cloudvsdatacenter}, \sys significantly outperforms DistServe and \jyh{vLLM}, achieving up to 2.5$\times$ and on average 1.8$\times$ lower E2E latency deadlines. This advantage stems from \sys's ability to deploy 3$\times$ more model replicas on the cloud than DistServe on the in-house server within the same price budget, which provides \sys with superior parallel processing capability. And the scheduling algorithm of \sys takes full advantage of the heterogeneity of cloud GPUs and improves system performance by allocating appropriate GPUs for prefilling and decoding, respectively. Thus, \sys greatly improves the cost-efficiency of deploying LLM services on the cloud.


\noindent \textbf{System throughput comparisons.} 
We further compare the system throughput between \sys and the baselines. 
As demonstrated in \autoref{fig:throughput}, compared with HexGen, \sys achieves 1.5 and 1.3$\times$ higher throughputs in coding and conversation workloads respectively. And compared with DistServe in the in-house setting, 1.5 and 2.1$\times$ higher throughputs are realized. These results demonstrate \sys's ability to effectively manage larger loads.


\begin{figure*}[!t]
    \centering
    \begin{minipage}{0.25\linewidth}
        \includegraphics[width=\linewidth]{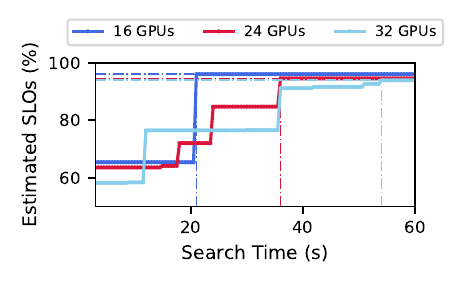}
          \vspace{-2em}
        \caption{\small{Convergence curves of scheduling from scratch for different cluster sizes.}}
          \vspace{-1em}
        \label{fig:schedulingoverhead}
    \end{minipage} \hspace{2pt}
    \begin{minipage}{0.35\linewidth}
        \includegraphics[width=\linewidth]{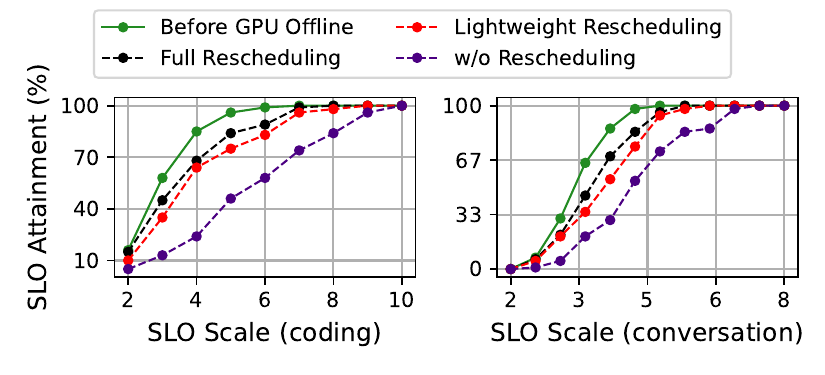}
          \vspace{-2em}
        \caption{\small{SLO attainments of before (indicated by the solid line) and after (indicated by the dotted lines) 4 out of 32 GPUs offline.}}
          \vspace{-1em}
        \label{fig:rescheduling}
    \end{minipage} \hspace{2pt}
    \begin{minipage}{0.35\linewidth}
        \includegraphics[width=\linewidth]{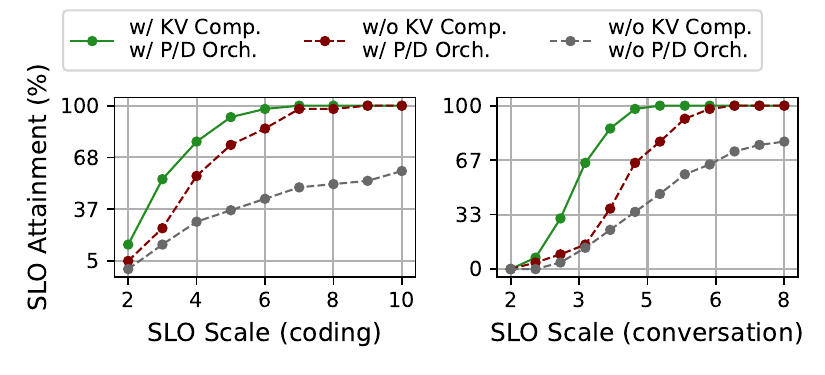}
          \vspace{-2em}
        \caption{\small{Impact of KV cache compression and prefill and decode orchestration on the SLO attainments.}}
          \vspace{-1em}
        \label{fig:compress}
    \end{minipage}
\end{figure*}

\begin{table}[!t]
\centering
\small
\caption{\small{Model deployment discovered by \sys.}}
\resizebox{\linewidth}{!}{
\begin{tabular}{c|l|l|l}
\hline
\textbf{Workload} & \textbf{GPU Configuration} & \textbf{Strategy} & \textbf{Type of Replicas} \\
\hline
\multirow{7}*{\rotatebox[origin=c]{90}{{Coding}}}
 & 8$\times$A40 & TP=2, PP=1 & 4 Prefill Replicas \\
 & 4$\times$A5000 & TP=4, PP=1 & 1 Prefill Replica \\
 & 4$\times$A6000 & TP=2, PP=1 & 2 Prefill Replicas \\
 & 2$\times$A5000+2$\times$3090Ti & TP=2, PP=2 & 1 Prefill Replica \\
 & 4$\times$3090Ti & TP=2, PP=2 & 1 Decode Replica \\
 & 4$\times$A6000 & TP=1, PP=2 & 2 Decode Replicas \\
 & 2$\times$A5000+2$\times$3090Ti & TP=2, PP=2 & 1 Decode Replica \\
\hline
\multirow{7}*{\rotatebox[origin=c]{90}{{Conversation}}}
 & 6$\times$A40 & TP=2, PP=1 & 3 Prefill Replicas \\
 & 2$\times$A5000+2$\times$3090Ti & TP=2, PP=2 & 1 Prefill Replica \\
 & 4$\times$3090Ti & TP=2, PP=2 & 1 Decode Replica \\
 & 2$\times$A40 & TP=1, PP=2 & 1 Decode Replica \\
 & 4$\times$A5000 & TP=2, PP=2 & 1 Decode Replica \\
 & 8$\times$A6000 & TP=1, PP=2 & 4 Decode Replicas \\
 & 2$\times$A5000+2$\times$3090Ti & TP=2, PP=2 & 1 Decode Replica \\
\hline
\end{tabular}
}
\label{table:gpu_config_strategy}
  \vspace{-2em}
\end{table}

\subsection{Case Study of Scheduling}
\label{sec:casestudyofscheduling}
\ffc{Table~\ref{table:gpu_config_strategy} presents the model deployment discovered by our scheduling algorithm. It can be seen that \sys prioritizes the GPUs with better computing ability for prefilling (e.g., A40) and those with higher memory access bandwidth GPUs for decoding (e.g., 3090Ti). 
Additionally, \sys automatically assigns more decode replicas to the conversation workload due to its longer output lengths, while more prefill replicas to the coding workload due to the longer prompt lengths. 
These results verify that \sys is able to take the heterogeneity in both hardwares and serving workloads into account, generating model deployment that maximizes the system performance.}

\ffc{Compared to the in-house setting with an 8$\times$A100 instance, where only 4 model replicas can be served, \sys serves a maximum of 12 model replicas in the cloud setting within the same price budget. Although individual inference processes in the cloud setting may experience increased latency due to the lower hardware performance (i.e., the cloud GPUs are less performant than A100), the overall performance is improved due to the higher number of model replicas. Thus, \sys conveys the ability of high-performance and cost-efficient LLM serving on clouds.}

%

\subsection{Effectiveness and Ablation Studies}
\label{sec:effectiveness_and_ablation}
\noindent \textbf{Time cost of Scheduling.}
We evaluate the running time of our scheduling algorithm from scratch with cluster sizes of 16, 24 and 32 GPUs.
Leveraging our effectively designed neighborhood construction method, the algorithm based on tabu search scales well with the number of GPUs, requiring approximately 21, 36 and 54 seconds to converge, as shown in \autoref{fig:schedulingoverhead}. This search process is executed once before the initial deployment of the system, rendering its time cost negligible given the hourly scale of online services. 




\noindent \textbf{Lightweight re-scheduling.}
To evaluate the effectiveness of our lightweight re-scheduling, we consider a scenario where 4 out of 32 GPUs become unavailable. Specifically, we remove two decode replicas and let \sys re-schedule on the fly. 
We compare our lightweight re-scheduling with \mytextcircled{1} a full re-scheduling approach, which involves re-starting the services and reloading parameters, and \mytextcircled{2} a no re-scheduling approach, which does not make any changes to the deployment plan and keeps the services using the remaining GPUs. 
As shown in \autoref{fig:rescheduling}, our lightweight re-scheduling achieves similar SLO attainment to the full re-scheduling approach and outperforms the no re-scheduling approach, showing the strengths of our lightweight re-scheduling. 
More importantly, our lightweight re-scheduling process finishes within seconds, without any overhead on parameter reloading, far exceeding the full re-scheduling approach. This is reasonable as tabu search is done locally and we only adjust the phase designation and orchestration to adapt to different cluster sizes. 
Therefore, \sys is able to handle the dynamicity in cloud environments well.




\begin{table}[!t]
\centering
\small
\caption{\jyh{\small{Overhead of Full and Lightweight Rescheduling.}}}
\jyh{
\begin{tabular}{l c c c}
\hline
\textbf{} & \textbf{Rescheduling} & \textbf{Reloading} & \textbf{Overall}\\ \hline
Full & 54($\pm$5)s & 103($\pm$10)s & 157($\pm$13)s \\ \hline
Lightweight & 13($\pm$2)s & 0s & 13($\pm$2)s \\ \hline
\end{tabular}
}
\label{tab:lr}
  \vspace{-2em}
\end{table}

\noindent \textbf{KV cache compression and orchestration.} 
There are two major techniques in \sys to address the communication cost of KV cache transmission from prefill to decode replicas, which are the KV cache compression and orchestration method. We conduct experiments to assess their effects. 
As illustrated in \autoref{fig:compress}, \sys exhibits a degraded performance in both coding and conversation scenarios without KV cache compression, incurring approximately 1.3$\times$ the overhead per single request. This undermines the benefits of phase splitting. If we further disable the orchestration method described in \S\ref{sec:schedule_lower_level} and substitute it with a random dispatching, there is another 4$\times$ of performance degradation. 
In summary, \sys chooses the parallel configurations and KV cache communication paths that optimize overall system performance given the high heterogeneity of communication bandwidth on the cloud.

\noindent \textbf{More Experiment Results.} 
We have also conducted more experiments to demonstrate the strengths of \sys, including more case studies, the effect of KV cache compression on model quality, and the accuracy of our simulator. Due to the space constraint, we refer interested readers to our supplemental material for more details.

\vspace{-1em}
\section{Conclusion}
\label{sec:con}
This paper explores the potential of deploying LLM services on clouds. Toward this end, we presented \sys, a system that employs hybrid model parallelism and phase splitting to enhance LLM serving efficiency across heterogeneous cloud GPU clusters.
With \sys, we proposed a novel scheduling algorithm that co-optimizes resource allocation, phase designation, parallelism strategies, and the orchestration of both prefill and decode phases. 
Additionally, we proposed a lightweight re-scheduling mechanism to enhance \sys performance in response to fluctuating online workloads for extremely fast adjustment on clouds. 
We conducted experiments on various workloads in both heterogeneous cloud and homogeneous in-house settings to demonstrate that \sys 
outperforms state-of-the-art systems within the same price budget.

\nocite{zhang2025sageattention,zhang2024sageattention2,zhang2025spargeattn,jiang2025demystifying}

\bibliography{citation}
\bibliographystyle{mlsys2025}


\appendix
\clearpage

\section{NP-Hardness of Deployment Planning}
\label{appendix:a}
Finding the optimal deployment plan that maximizes the overall SLO for deploying multiple models on a heterogeneous GPU cluster with variable interconnect topology and computational capabilities is non-trivial. In particular, we show that this problem is NP-hard by transforming it into the well-known NP-hard Job Shop Scheduling Problem (JSSP) \cite{sotskov1995np,omar2006job}. 

\noindent \textbf{Transformation of the deployment planning problem to JSSP.} Each GPU in the heterogeneous cluster serves as a distinct machine in the JSSP. These GPUs exhibit differences in computation power, memory, and communication capabilities. Each model, or its components depending on the placement method, is considered a job within JSSP. The deployment of each model involves multiple tasks or operations, each corresponding to the deployment of a part of a model on one or more GPUs, accompanied by specific resource requirements and execution constraints. Sequential dependencies are evident in scenarios where the completion of one operation on a GPU is prerequisite for the initiation of the next on another GPU, characteristic of pipeline model parallelism. Concurrent dependencies arise when operations must occasionally synchronize across GPUs, reflecting interdependencies that require coordination akin to those in tensor model parallelism. In this context, maximizing SLO does not solely involve minimizing idle and wait times but also necessitates the optimization of the allocation and scheduling of operations to ensure continuous and efficient GPU utilization. Thus, this challenge can be viewed as a variant of JSSP where the objective shifts from minimizing makespan to maximizing SLO, analogous to maximizing the number of completed jobs or operations within certain latency deadlines. This requires managing both the sequence and concurrency of operations across heterogeneous resources and optimizing overall system efficiency to mitigate bottlenecks and reduce synchronization overheads.

Job shop scheduling is recognized as NP-hard due to the complexity inherent in managing dependencies and varying capabilities across machines. By formulating this problem as a variant of JSSP adapted for SLO, we establish that solving the model placement problem is at least as hard as solving the classic NP-hard JSSP, thus confirming the NP-hardness of the problem.

\begin{algorithm}[!t]
\caption{Generate Model Parallel Configurations}
\small
\label{alg:genmodelparallel}
\begin{algorithmic}[1]
\STATE \textbf{Initialize:} group formation: $G = \{G_1, G_2, ..., G_g\}$, minimum number of single-type GPUs in the group: $T = \{T_1, T_2, ..., T_g\}$, cluster information: $I$, model configuration: $M$ 
\STATE $model\_parallel\_configurations \leftarrow []$
\FOR{i in len($G$)}
\STATE $plan\_list \leftarrow []$
\STATE{\textcolor{gray}{/* Limit TP within Single-type GPUs */}}
\FOR{$TP$ in $\{1, 2, ..., T_i\}$}
        \FOR{$PP$ in $\{1, 2, ..., \frac{G_i.num\_gpus}{T_i}\}$}
            \STATE{\textcolor{gray}{/* Route Pipeline Communication */}}
            \STATE $plan \gets \texttt{Dynamic\_Programming}(I,TP,PP)$
            \STATE{\textcolor{gray}{/* Generate Pipeline Partition */}}
            \STATE $plan \gets \texttt{Pipeline\_Partition}(M, plan)$
                \IF{$G_i.type$ is prefill}
                \STATE $C \leftarrow \text{latency}(plan)$
            \ELSE
                \STATE $C \leftarrow \text{throughput}(plan)$
            \ENDIF
            \STATE $plan\_list.append((C, plan))$
        \ENDFOR
    \ENDFOR
    \IF{$G_i.type$ is prefill}
        \STATE{\textcolor{gray}{/* Select Latency Optimal Plan */}}
        \STATE $plan \leftarrow \min(C)$ in $plan\_list$
    \ELSE
        \STATE{\textcolor{gray}{/* Select Throughput Optimal Plan */}}
        \STATE $plan \leftarrow \max(C)$ in $plan\_list$
    \ENDIF
    \STATE $model\_parallel\_configurations.append(plan)$
\ENDFOR
\STATE \textbf{return} $model\_parallel\_configurations$
\end{algorithmic}
\end{algorithm}

\section{Deduction of Parallel Configuration}
\label{appendix:b}
Given the group formation and the designated phase, we need to deduce the optimal parallel configuration for each group. Algorithm \autoref{alg:genmodelparallel} outlines the process. \mytextcircled{1} We enumerate all possible TP and PP combinations on each given group formation. Note that our first heuristic is to limit tensor model parallelism within single-type GPUs, so the TP degree should be smaller or equal to the minimum number of single-type GPUs in the group, which largely minimizes the search space. \mytextcircled{2} Dynamic programming algorithm is utilized to route the pipeline communication path. It optimizes communication routing in a network by using a bitmask to represent all possible subsets of stages, initializes each stage with a zero bandwidth and builds paths by calculating the potential bandwidth for each link between stages, updates the optimal path recursively if a higher bandwidth stage is found, and determines the maximum bandwidth path available by examining the states for the subset that includes all stages, ensuring the most efficient data transfer across the network.  \mytextcircled{3} We adjust the pipeline layer partition with respect to the memory capacity and computing ability of different GPU types. Specifically, the pipeline partition is adjusted in proportion to the total memory and computing capacity of the GPU set currently servicing this stage, while ensuring that the memory limits of individual GPUs are not exceeded. This heuristic has proven effective in determining an optimal pipeline partition. \mytextcircled{4} For the compute-bound prefill replicas, we select the latency optimal plans, for the memory bandwidth-bound decode replicas, we select the throughput optimal plans. To estimate the latency and throughput of each plan, we employ the cost model proposed by HexGen \cite{jiang2024hexgen}, which directly provides us with the inference memory and latency costs for both prefill and decode phases, relative to different request batch sizes. We calculate the throughput by dividing the maximum total batched token size that the device group can handle by the decode latency. Note that the estimated latency information is also provided to our simulator for SLO estimation.



\section{Inter-connection Bandwidth Matrix}
\label{appendix:comm_matrix}
\begin{figure}[!t] 
  \centering
  \includegraphics[width=\linewidth]{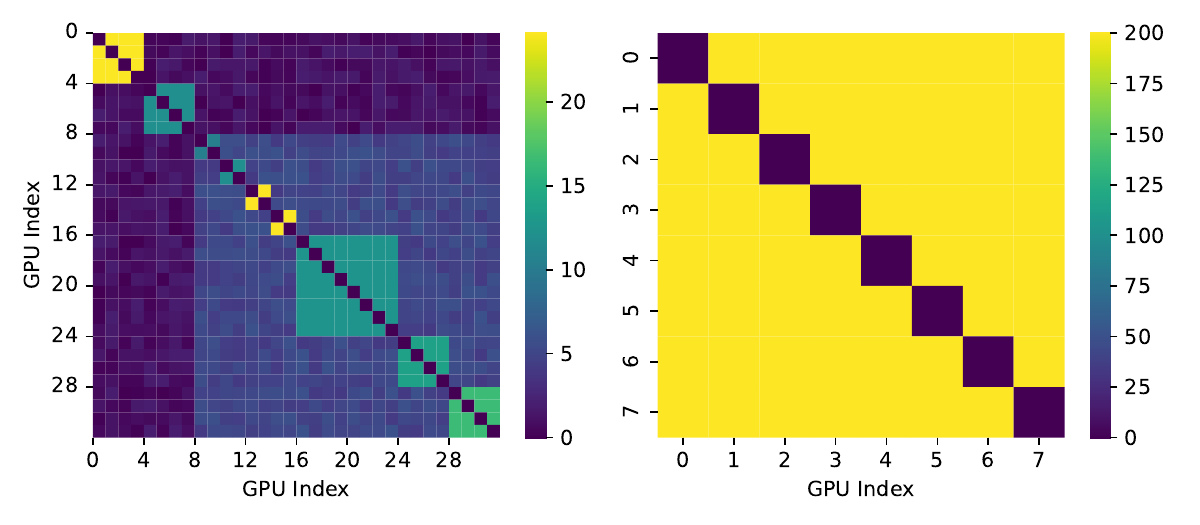} 
  \caption{\jyh{Heat map of inter-connection bandwidth matrix in the cloud (left) and in-house (right) settings.}}
  \label{fig:comm_matrix}
\end{figure}

The bandwidth distributions exhibit significant variability in cloud and in-house environments. We measure the communication bandwidth between each pair of GPUs via NCCL for both environments described in \S\ref{sec:eva_setup}. As shown in the left heatmap of \autoref{fig:comm_matrix}, the cloud environment demonstrates notable bandwidth heterogeneity, influenced by a range of GPU types and network configurations. This variability results in non-uniform connectivity patterns across the network. Conversely, the right heatmap showcases the in-house environment, characterized by a uniform GPU-to-GPU communication bandwidth, evidenced by consistently high connectivity values. These visualizations emphasize the distinctions between cloud and in-house environments.

\section{Ratio Impact on System SLO Attainment}
\label{appendix:ratio}

\begin{figure}[!t] 
  \centering
  \includegraphics[width=\linewidth]{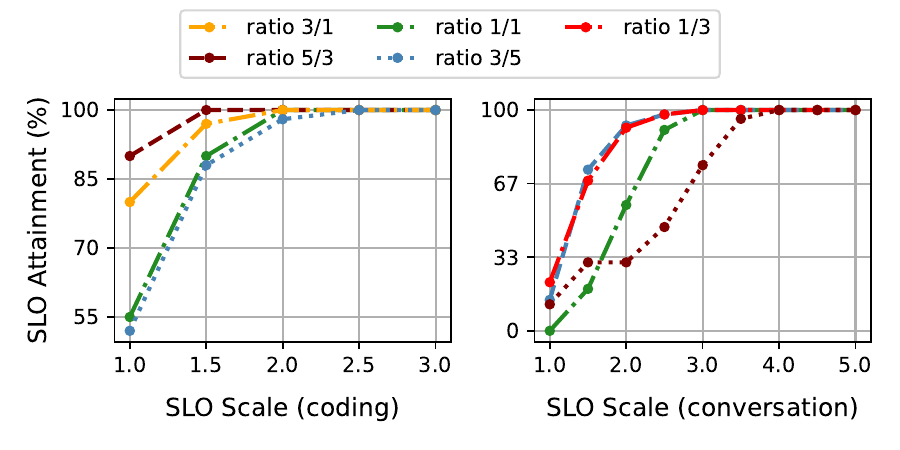} 
  \caption{{Impact of phase designation and orchestration on overall system \jyhh{SLO attainment}. We experiment with LLaMA-13B on both coding and conversation workloads across 16 A5000 GPUs, with two GPUs serving one replica.}}
  \label{fig:ratio_goodput}
\end{figure}

We show the impact of phase designation and orchestration on overall system SLO attainment in~\autoref{fig:ratio_goodput}.
The coding workload, characterized by relatively longer input length and
shorter output length, exhibits enhanced performance with more prefill replicas and fewer decode replicas. A ratio of 5:3 yields the optimal results. Conversely, the conversation workload, typified by relatively shorter prompts and longer responses, necessitates more decode replicas and fewer prefill replicas to prioritize resources to the long-running decoding. Here, a ratio of 3:5 achieves the best performance.

\section{Implementation Details}
\label{appendix:components}

\begin{figure}[!t]
  \centering
  \includegraphics[width=\linewidth]{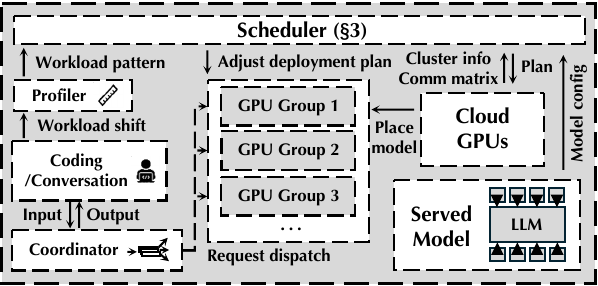} 
  \vspace{-1em}
  \caption{System overview of \sys.}
  \label{fig:sysoverview}
  \vspace{-1em}
\end{figure}

\noindent \textbf{Overview of \sys.} 
The architecture overview of \sys is shown in \autoref{fig:sysoverview}. There are three major components, which are the scheduler, the workload profiler, and the task coordinator.

The \textit{scheduler} is the core of \sys for high-performance LLM serving in cloud environments.  
The scheduler takes as input the model configurations (e.g., hidden size and layer number), workload patterns obtained from the workload profiler, cluster information (e.g., available GPUs and their corresponding types), and communication bandwidth matrix among all GPUs. Then, it performs the scheduling algorithm introduced in \S\ref{sec:scheduling} to provide the optimal deployment plan. 
Should there be a detected shift in workload, or a GPU heartbeat timeout that suggests a need for cluster size adjustment, the scheduler will perform the lightweight re-scheduling process and adjust the deployment plan to adapt to the new workload or cluster size. 

The \textit{workload profiler} monitors the real-time workload patterns, including the average prompt length of incoming requests and average output length of generated responses. These patterns are utilized to analyze the prefill and decode cost for each single request. 
For instance,  in contemporary LLM services, common workload scenarios include coding and conversation \cite{patel2023splitwise}, where both typically have a median prompt length exceeding 1000 tokens. However, the coding service produces much fewer output tokens, with a median of 13, while the conversation service generates a larger number of output tokens, with a median of 129. Undoubtedly, the overall system workload varies when the proportions of incoming requests for various services change in real-time.
Once an obvious workload shift is detected, the workload profiler will notify the scheduler.

The \textit{task coordinator} is in charge of the request dispatching among the prefill and decode replicas. 
Upon receiving a request, the task coordinator assigns the appropriate prefill replica and decode replica, respectively. The assignment is guided by the deployment plan generated by the scheduler.
The task coordinator is mainly based on an open-source implementation of decentralized computation coordination \cite{yao2023open} that utilizes libP2P \cite{libp2p} to establish connections among the work groups in a peer-to-peer network.

Based on these components, the overall routine of \sys is as follows.
\mytextcircled{1} To launch a serving process, the scheduler generates the deployment plan, which is then utilized to instantiate the model replicas over the cloud GPU resources. 
\mytextcircled{2} During the serving process, the coordinator dispatches the incoming requests across the prefill and decode replicas, and gathers the generated responses. 
\mytextcircled{3} At the same time, the workload profiler consistently monitors the workload and reports to the scheduler. 
\mytextcircled{4} Once a workload shift is detected, the scheduler triggers a lightweight re-scheduling process to adjust the deployment plan for better adaptation to the new workload.

\textbf{Parallel communication groups.} All communication primitives in \sys are implemented using NVIDIA Collective Communication Library (NCCL). To circumvent the substantial overhead associated with constructing NCCL groups, \sys preemptively establishes a global communication group pool containing all potentially required groups. For KV cache communication, we employ NCCL's asynchronous \texttt{SendRecv}/\texttt{CudaMemcpy} functions for KV cache communication to prevent GPU blocking and enable computation and communication overlapping during transmission. KV cache queues are maintained on the prefill replicas, and upon completion of a decoding round, the decode replicas retrieve KV caches from these queues, utilizing the GPU memory of the prefill replicas as queuing buffers.

\section{Case Study of Scheduling}
\label{appendix:scheduling_results}
We list the deployment plan generated by \sys from coding workload to conversation workload in the heterogeneous setting. We use the following representation to describe the scheduled results. We use an array to specify one independent model replica, with two numbers representing the degrees of tensor model parallelism and pipeline model parallelism. For example, (2,2) indicates a model replica with tensor model parallel degree of 2 and pipeline model parallel degree of 2 (2 pipeline stages).

We also provide the instances we considered in \S\ref{sec:eva_setup} here for better readability: two 4$\times$A6000 instances, \jyh{two 4$\times$A5000 instances}, one 8$\times$A40 instance and two 4$\times$3090Ti instances, making up to be 32 GPUs in total.

\noindent \textbf{Parallel configuration breakdown.} In the coding workload, the 8$\times$A40 instance employs a parallel strategy (2,1) to support four prefill replicas. One 4$\times$A6000 instance uses a parallel strategy (2,1) to support two prefill replicas, while the other one 4$\times$A6000 instance uses a parallel strategy (1,2) for two decode replicas. 
\jyh{One 2$\times$A5000 and one 2$\times$3090Ti instances utilize a parallel strategy (2,2) to support one prefill replica, and the other one 2$\times$A5000 and one 2$\times$3090Ti instances utilize a parallel strategy (2,2) to support one decode replica. One 4$\times$A5000 instance utilizes a parallel strategy (4,1) to support one prefill replica. One 4$\times$3090Ti instance implements a parallel strategy (2,2) to support one decode replica.}

In the conversation workload, the 8$\times$A40 instance employs parallel strategies (2,1) and (1,2) to support three prefill replicas and one decode replica, respectively. The two 4$\times$A6000 instances utilize a parallel strategy (1,2) to support four decode replicas. 
\jyh{One 2$\times$A5000 and one 2$\times$3090Ti instances utilize a parallel strategy (2,2) to support one prefill replica, and the other one 2$\times$A5000 and one 2$\times$3090Ti instances utilize a parallel strategy (2,2) to support one decode replica. One 4$\times$A5000 instance utilizes a parallel strategy (2,2) to support one decode replica. One 4$\times$3090Ti instance implements a parallel strategy (2,2) to support one decode replica.}

\noindent \textbf{Insights.} In the in-house setting, the 8$\times$A100 instance can only serve 4 model replicas, while in the cloud setting, the 32 cloud GPUs with various types can serve a maximum of 12 model replicas with various parallel configuration within the same price budget. In this case, although individual inference tasks in the cloud setting may experience increased latency due to the lower hardware performance (e.g., GPU flops and bandwidth), the overall system performance is improved due to the higher number of model replicas. Additionally, our scheduling algorithm prioritizes GPUs with high peak fp16 flops for prefilling (e.g., A40) and high memory bandwidth GPUs for decoding (e.g., 3090Ti), and selects the most suitable model parallel configuration for each phase to optimize the overall system performance. \jyh{And although KV cache compression can linearly mitigate communication overhead, significant disparities in bandwidth across different cloud environments render extremely low bandwidth scenarios—such as those experienced between data centers—unsuitable for effective KV cache communication. Thanks to our scheduling and orchestration algorithms, \sys automatically identifies KV cache transmission paths that maintain overall performance.}

\section{Case Study of Lightweight Rescheduling}
\label{appendix:lightweight_rescheduling}
We list the deployment plan change during lightweight rescheduling with 4 out of 32 GPUs (one 4$\times$A6000 instance that support two decode replicas) become unavailable.

The deployment plan for the coding workload, detailed in Appendix \ref{appendix:scheduling_results}, initially includes 8 prefill and 4 decode replicas. After the offline of 4 GPUs, there are 8 prefill and 2 decode replicas remaining. A subsequent lightweight rescheduling converts one prefill replica, which uses 4 A5000 GPUs with a (4,1) strategy, into a decode replica. The adjustment is reasonable as this group of GPUs exhibits the highest overall memory bandwidth among the prefill replicas. 
The deployment plan for the conversation workload initially includes 4 prefill and 8 decode replicas. After the offline of 4 GPUs, there are 4 prefill and 6 decode replicas remaining. A subsequent lightweight rescheduling converts one prefill replica, which uses 2 A40 GPUs with a (2,1) strategy, into a decode replica.

\jyh{
\section{Case Study of Network Effect on Phase Splitting}
\label{appendix:networkeffect}

\begin{figure}[!t] 
  \centering
  \includegraphics[width=\linewidth]{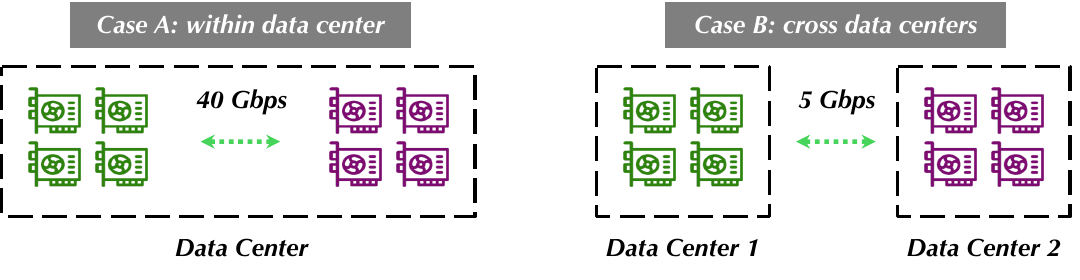} 
  \caption{\jyh{Two exampled network conditions on cloud.}}
  \label{fig:caseab}
\end{figure}

\begin{figure}[!t] 
  \centering
  \includegraphics[width=\linewidth]{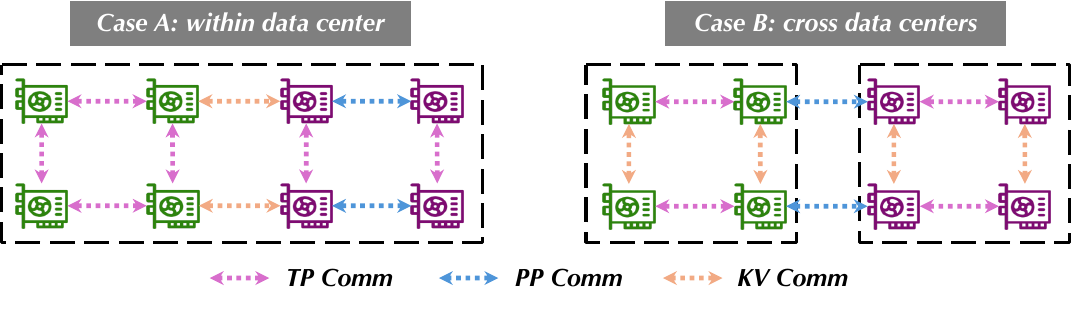} 
  \caption{\jyh{\sys deployment plans on different cases.}}
  \label{fig:caseab2}
\end{figure}

\begin{table}[h]
\centering
\jyh{
\small
\caption{\jyh{Benchmarks of non-disaggregation baseline vs. \sys under high inter-instance communication bandwidth vs. \sys under low inter-instance communication bandwidth.}}
\resizebox{\linewidth}{!}{
\begin{tabular}{ l c c c c }
\hline
\textbf{Configuration} & \textbf{Prefill} & \textbf{KV Comm} & \textbf{Decode} & \textbf{E2E Throughput} \\
\hline
Baseline & 884 ms & 0 ms & 1689 ms & 1610 tokens/s\\
\hline
\sys (High) & 698 ms & 133 ms & 1126 ms & 3292 tokens/s \\
\hline
\sys (Low) & 964 ms & 41 ms & 1846 ms & 2196 tokens/s \\
\hline
\end{tabular}
\label{tab:dandnod}
}
}
\end{table}


\begin{table}[!t]
\centering
\small
\caption{\jyh{Impact of KV cache communication compression on the perplexity results on WikiText2, PTB and CBT datasets.}}
\jyh{
\begin{tabular}{c c c c}
\hline
\textbf{Dataset} &  & \textbf{LLaMA-7B} & \textbf{LLaMA-30B} \\ \hline
\multirow{2}{*}{WikiText2} & 16-bit & 3.53 & 2.73 \\ 
                                    & 4-bit  & 3.55 & 2.75 \\ \hline
\multirow{2}{*}{PTB} & 16-bit & 7.46 & 6.49 \\ 
                                       & 4-bit  & 7.42 & 6.55 \\ \hline
\multirow{2}{*}{CBT} & 16-bit & 7.66 & 6.31 \\ 
                                               & 4-bit  & 7.70  & 6.30  \\ \hline
\end{tabular}
}
\label{tab:PPL}
\end{table}

\begin{table}[!t]
\centering
\small
\caption{\jyh{LLaMA rouge results (using 16-bit outputs as the ground truth and the 4-bit outputs as the prediction) on WikiText2, PTB and CBT datasets.}}
\jyh{
\begin{tabular}{c c c c}
\hline
\textbf{Dataset} &  & \textbf{LLaMA-7B} & \textbf{LLaMA-30B} \\ \hline
\multirow{3}{*}{WikiText2} & ROUGE-1 & 0.962 & 0.942 \\ 
                           & ROUGE-2 & 0.941 & 0.928 \\ 
                           
                           & ROUGE-L & 0.955 & 0.941 \\ \hline
\multirow{3}{*}{PTB}       & ROUGE-1 & 0.975 & 0.928 \\ 
                           & ROUGE-2 & 0.950  & 0.911 \\ 
                           & ROUGE-L & 0.971 & 0.928 \\ \hline
\multirow{3}{*}{CBT}       & ROUGE-1 & 0.925 & 0.946 \\ 
                           & ROUGE-2 & 0.912 & 0.931 \\ 
                           & ROUGE-L & 0.925 & 0.937 \\ \hline
\end{tabular}
}
\label{tab:rouge_performance}
\end{table}

\begin{table}[h]
\centering
\jyh{
\small
\caption{\jyh{Benchmarks of \sys with 16-bit vs. 4-bit communications.}}
\resizebox{\linewidth}{!}{
\begin{tabular}{ l c c c c }
\hline
\textbf{Configuration} & \textbf{Prefill} & \textbf{KV Comm} & \textbf{Decode} & \textbf{E2E Throughput} \\
\hline
16-bit & 684 ms & 584 ms & 1108 ms & 2450 tokens/s \\
\hline
4-bit & 698 ms & 133 ms & 1126 ms & 3292 tokens/s \\
\hline
\end{tabular}
\label{tab:add}
}
}
\end{table}


Consider use \sys to serve LLaMA-30B model in a heterogeneous environment featuring two GPU instances: the first instance equipped with 4$\times$A40 GPUs, and the second with 4$\times$3090Ti GPUs. We conducted tests on the inference throughput of this setup by feeding it continuous input sequences of length 1024 under two different inter-instance communication bandwidths: 40 Gbps and 5 Gbps, as demonstrated in \autoref{fig:caseab}.

We established a non-disaggregating baseline that utilizes 4$\times$A40 GPUs to support one model replica and 4$\times$3090Ti GPUs to support another. By comparing the baseline with \sys under different network conditions, we observed some interesting results: With a bandwidth of 40 Gbps, \sys leverages the 4$\times$A40 GPUs with higher peak flops to support one prefill replica, and the 4$\times$3090Ti GPUs with higher memory access bandwidth to support one decode replica. This configuration optimizes system performance, achieving a 2$\times$ performance gain over the non-disaggregating baseline. However, at a lower bandwidth of 5 Gbps, the inter-instance communication bandwidth is insufficient for efficient KV cache communication. Consequently, \sys allocates 2$\times$A40 GPUs and 2$\times$3090Ti GPUs to both prefill and decode replica, which utilizes intra-instance high network bandwidth for KV cache communication and inter-instance low network bandwidth for pipeline communication, resulting in a 1.4$\times$ improvement over the non-disaggregating baseline. The illustration of deployment plans are demonstrated in \autoref{fig:caseab2}, the single request prefill/decode/KV cache communication time and overall system throughputs are demonstrated in \autoref{tab:dandnod}.
}

\begin{figure}[!t]
  \centering
  \includegraphics[width=\linewidth]{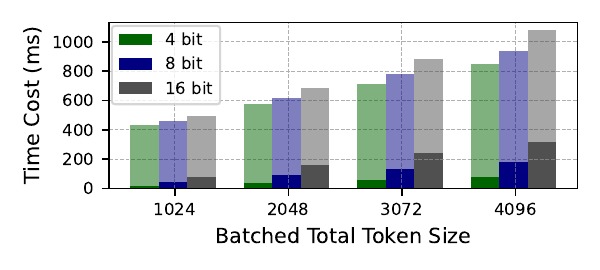}
    \vspace{-1em}
  \caption{\jyh{Impact of KV cache communication compression. (Non-transparent: time cost of KV cache communication. Transparent: end-to-end processing time.)}}
  \label{fig:kvcache}
    \vspace{-1em}
\end{figure}

\jyh{
\section{PPL and ROUGE results on KV cache compression}
\label{appendix:ppl}
We list the PPL and ROUGE results of LLaMA-7B and LLaMA-30B models on WikiText2, PTB and CBT datasets with both 16-bit and 4-bit KV cache precision levels, as shown in \autoref{tab:PPL} and \autoref{tab:rouge_performance}. Experimental results have demonstrated that the PPL between 16-bit precision and 4-bit precision is within 1\% across all experimental scenarios, and the ROUGE-1, ROUGE-2 and ROUGE-L scores are around 0.95 across all cases, which confirms the validity of our approach. We also demonstrate the the benchmarks of \sys with 16-bit vs. 4-bit communications in \autoref{tab:add} with the same experimental setups as mentioned in \autoref{appendix:networkeffect}, and benchmarks in \autoref{fig:kvcache}, with two A5000 GPUs serving a LLaMA-7B model.
}

\begin{figure}[!t] 
  \centering
  \includegraphics[width=\linewidth]{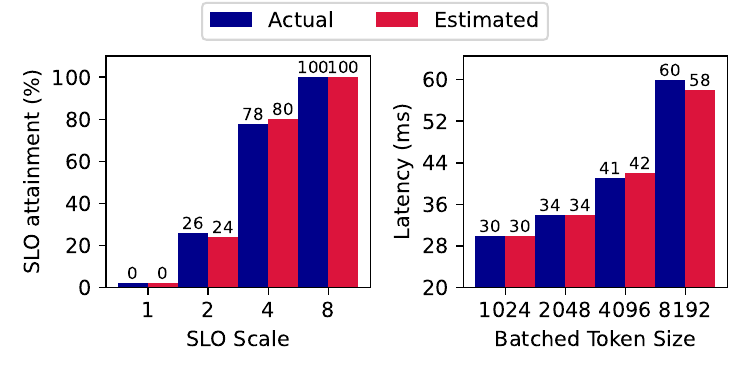} 
  \caption{\jyhh{Comparison of benchmarked and estimated performance metrics for simulator (left) and alpha-beta model (right).}}
  \label{fig:accuracy}
\end{figure}

\jyhh{
\section{Simulator and alpha-beta model accuracy}
\label{appendix:simu}
To assess the accuracy of the simulator and alpha-beta model for KV cache communication, we conducted a series of micro-benchmarks using the LLaMA-30B model. These benchmarks varied in SLO scales and batched token sizes to evaluate our estimation outputs against actual execution metrics, specifically SLO attainment and latency. The results, detailed in \autoref{fig:accuracy}, indicate that the simulator and alpha-beta model closely correspond with actual execution performance.
}


\end{document}